\documentclass[letterpaper]{article}
\usepackage{natbib,alifeconf}
\usepackage{xcolor}
\usepackage{amsmath}
\usepackage{amssymb}

\newcommand{\norm}[1]{\left\lVert#1\right\rVert}

\def\RR{\mathbb{R}}

\title{A Quantification Approach for Transferability in Lifelike Computing Systems}
\author{Martin Goller$^{1}$ \and Sven Tomforde$^1$ \\%, Second Author$^2$
\mbox{}\\
$^1$Intelligent Systems, Christian-Albrechts-Universität zu Kiel, Germany \\
goller.cau@gmail.com / st@informatik.uni-kiel.de}

\begin{document}
\maketitle

\begin{abstract}
The basic idea of lifelike computing systems is the transfer of concepts in living systems to technical use that goes even beyond existing concepts of self-adaptation and self-organisation (SASO). As a result, these systems become even more autonomous and changeable - up to a runtime transfer of the actual target function. Maintaining controllability requires a complete and dynamic (self-)quantification of the system behaviour with regard to aspects of SASO but also, in particular, lifelike properties. In this article, we discuss possible approaches for such metrics and establish a first metric for transferability. We analyse the behaviour of the metric using example applications and show that it is suitable for describing the system's behaviour at runtime.
\end{abstract}

%%%%%%%%%%%%%%%%%%%%%%%%%%%%%%%%%%%%%%%%%%%%%%%%%%%%%%%%%%%%%%%%%%%%%%%
%
%
%       Sec 1: Intro 
%
%
%%%%%%%%%%%%%%%%%%%%%%%%%%%%%%%%%%%%%%%%%%%%%%%%%%%%%%%%%%%%%%%%%%%%%%%

\section{I. Introduction}

% SASO
The last two decades have seen a trend towards more autonomous behaviour in technical systems [\cite{zhang2017current}]. On the one hand, this leads to increased controllability by the human administrator, as self-adaptation of behaviour and self-organisation of the system structure occurs through the technical systems themselves (here summarised as SASO capabilities). On the other hand, this leads to even more complex solutions, as these autonomous SASO capabilities have to be developed, monitored and controlled [\cite{serugendo2011self}]. Examples can be found in the efforts done by initiatives such as Organic [\cite{MST17}] and Autonomic Computing [\cite{KC03}].

Despite this increasing usability and strength of SASO technologies, we continue to face major challenges: Systems are increasingly interconnected, they need to interact with unknown other systems, are used for novel problems or are customised at runtime. All this leads to a further increase in complexity and severely compromised controllability. The vision of lifelike computing systems [\cite{stein2021lifelike}] is to meet these challenges through further autonomy -- principles of living systems are transferred to technical use so that systems can change autonomously and take on new tasks.

In addition to the effects on system development and operation, questions of monitoring such system behaviour are increasingly coming to the fore. Starting from a quantification of aspects of the behaviour of lifelike systems, the focus of interest is on the self-explanation of behaviour as a fundamental aspect of future lifelike computing systems. 

In previous work, we have presented a quantification framework for runtime monitoring of SASO systems that enables statements about the associated SASO behaviour in addition to classical performance metrics. We propose to use this as a basis for further research into a quantification framework for lifelike computing systems. In this article, we take a step in this direction and summarise the existing framework in relation to lifelike systems and present a new metric for transferability. This measures the extent to which a system has adapted to another problem domain.

The remainder of this article is organised as follows: Section~2 revisits the measurement framework for macro-level behaviour assessment in relation to lifelike computing systems. This includes the definition of a basic system model and the definitions of existing measures for adaptation behaviour. Afterwards, Section~3 introduces ideas towards an extended measurement framework for lifelike systems. Section~4 then presents the novel measure for transferability. The behaviour is analysed in several simulation scenarios and presented in Section~5. Finally, Section~6 summarises the article and gives an outlook on future work.

\section{II. A Measurement Framework for Macro-Level Behaviour Assessment of Lifelike Systems}

The basis of our approach to a quantification of system properties (and a basis for self-explanatory mechanisms) of lifelike technical systems is an integrated framework that builds upon standard performance-oriented measures. We model externally observable aspects related to adaptation and evolution behaviour and detect changes during operation. To this end, in this section, we first present our system model, which we currently assume - and which can form the basis for future lifelike systems. Using this system model, we then explain existing and potential approaches for quantifying system properties.

\subsection{System Model}

In this article, we make use of the notion from the field of Organic Computing [\cite{MST17}] and refer to a technical system $S$ as a collection $A$ of autonomous subsystems $a_i$ that are able to adapt their behaviour based on self-awareness of the internal and external conditions. We further assume that such a subsystem is an entity that interacts with other entities, i.e., other systems, including hardware, software, humans, and the physical world with its natural phenomena. These other entities are referred to as the \textit{environment} of the given system. The \textit{system boundary} is the common frontier between the system and its environment.

\begin{figure}[hbt!]
	\centering
	\includegraphics[width=1\linewidth]{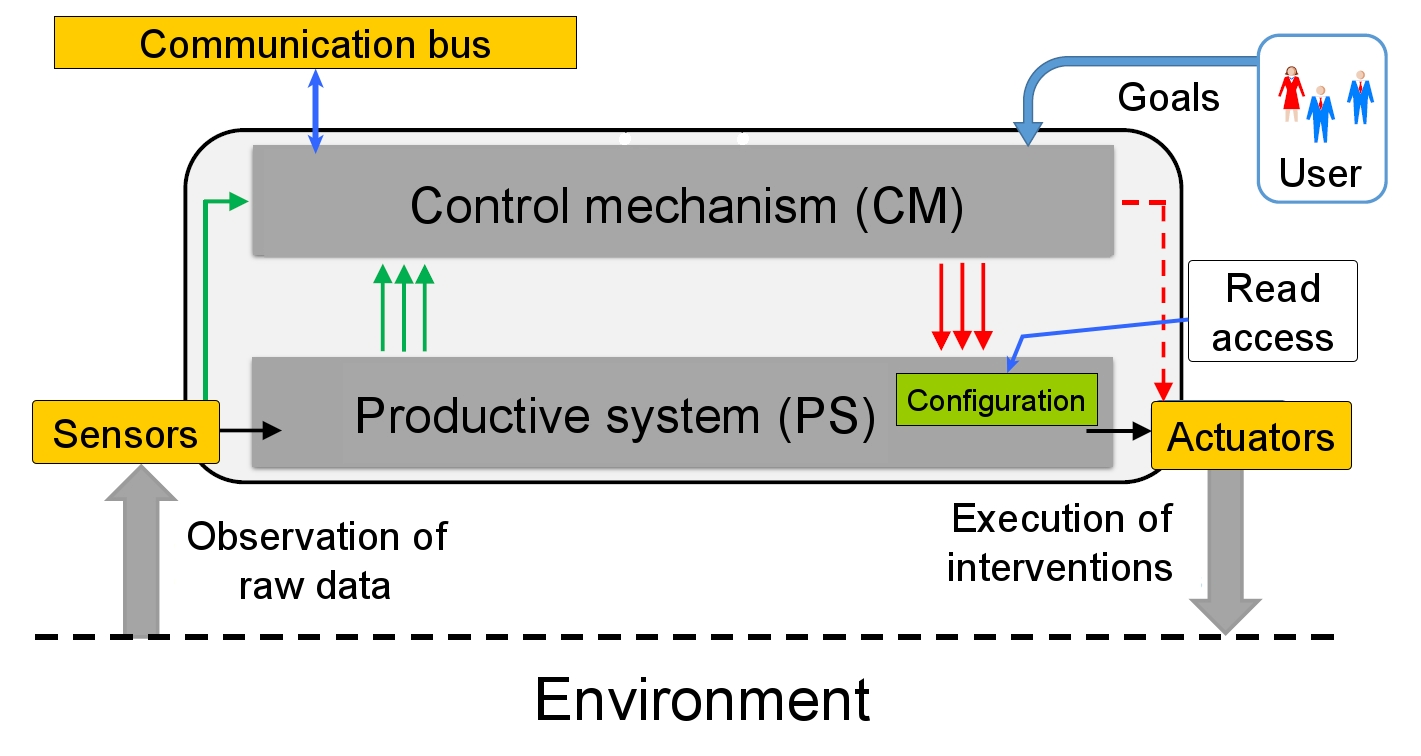}%{images/system-design.png}
	%{\small \caption{Schematic illustration of a subsystem $a_i$.}}
	\caption{Schematic illustration of a subsystem $a_i$ from~[\protect{\cite{TP+11}}]. The arrows from the sensors and PS to the CM indicate observation flows, while the arrows from the CM to the PS and the actuators indicate control flows. Dashed arrows emphasise a possible path that is typically not used. Not shown: The CM is able to communicate with other CMs in the shared environment to exchange information such as sensor reading and to negotiate policies using the communication bus.}\label{fig:SysPart}
\end{figure}

Each $a_i \in A$ is equipped with sensors and actuators (both, physical ad virtual). Internally, each $a_i$ consists of two parts: The productive system part \textit{PS}, which is responsible for the basic purpose of the system, and the control mechanism \textit{CM}, which controls the behaviour of the PS (i.e., performs self-adaptation) and decides about relations to other subsystems. In comparison to other system models, this corresponds to the separation of concerns between \textit{System under Observation and Control} (SuOC) and \textit{Observer/Controller} tandem [\cite{TP+11}] in the terminology of Organic Computing (OC)~[\cite{TomfordeSM17}] or \textit{Managed Resource} and \textit{Autonomic Manager} in terms of Autonomic Computing~[\cite{KC03}]. Figure~\ref{fig:SysPart} illustrates this concept with its input and output relations. The user describes the system purpose by providing a utility or goal function $U$ which determines the behaviour of the subsystem. The User usually takes no further action to influence the decisions of the subsystem. Actual decisions are taken by the productive system and the CM based on the external and internal conditions and messages exchanged with other subsystems. We model each subsystem to act \textit{autonomously}, i.e., there are no control hierarchies in the overall system. Please note that for the context of this article an explicit local configuration of the PS is necessary -- which in turn limits the scope of the applicability of the proposed methods. Furthermore, each subsystem must provide read-access to the configuration.

At each point in time, the productive system of each $a_i$ is configured using a vector $c_i$. This vector contains a specific value for each control variable that can be altered to steer the behaviour, independently of the particular realisation of the parameter (e.g., as real value, boolean/flag, integer or categorical variable). Each subsystem has its own configuration space, i.e. an n-dimensional space defining all possible realisations of the configuration vector. The combination of the current configuration vectors of all contained subsystems of the overall system $S$ defines the joint configuration of $S$. We assume that modifications of the configuration vectors are done by the different $CM$ only, i.e. locally at each subsystem, and are the result of the self-adaptation process of the $CM$.

This system model describes an approach based on the current state-of-the-art in the field of self-adaptive and self-organising systems. We assume that ongoing research towards more lifelike systems will shift the boundaries in terms of the underlying technology as well as the possibility to alter higher-levelled design decisions -- but it will most likely not result in entirely new design concepts. In turn, we assume that fundamental questions will arise about how the CM evolves according to the characteristics of its 'environmental niche', for instance, but the separation of concerns between CM and PS remain visible.

\subsection{Standard Measures}

Usually, the first aspects to measure when quantifying the behaviour of a technical system is the primary purpose of the system and its performance while working towards it.

Based on the categorisation proposed by McGeoch in [\cite{mcgeoch2012guide}], we can distinguish between the two performance aspects \textit{quality of the solution} and \textit{time required for the solution}. While the first measure makes a statement about how good the solution for the current task is it does not say anything about the time taken. Vice versa the latter measure says nothing about the quality of the solution.
 
Furthermore, these two aspects of system behaviour can be examined with more theoretical approaches. For instance, the O(n) notation, runtime or memory complexity can be quantified. This can be extended with a verification of the underlying processes, i.e., guarantees that may be quantified in terms of coverage or degree of guarantee-able behaviour. As an alternative, the 'restore-invariant approach' by Nafz et al.~[\cite{nafz2011constraining}] establishes a formal framework for self-organisation behaviour that may serve as a quantifiable basis.

In addition to these considerations, the robustness and resilience of systems, as well as their behaviour, can be quantified using specific metrics. An example from the field of Organic Computing systems can be found in [\cite{TomfordeKMBS18}].

\subsection{Self-Adaptation based Measures}
Due to the separation of responsibility in SASO systems -- the CM takes autonomous decisions while the PS is responsible for their execution -- research on SASO systems entailed extending the measurement framework by several SASO-specific metrics. For instance, Kaddoum et al.~[\cite{Kad10}] discuss the need to refine classical performance metrics to SASO purposes and present specific metrics for self-adaptive systems. They distinguish between 'nominal' and 'self-*' situations and their relations: The approach determines the effort by comparing the operation time with the adaptation time. Some of the developed metrics have been investigated in detail by Camara et al. for software architecture scenarios~[\cite{Cam14}]. Besides, success and adaptation efforts and ways to measure autonomy have been investigated, see e.g.~[\cite{Gronau2016}].

In addition to these goal- and effort-based metrics, several other measurements examine the macro-level behaviour of a set of autonomous subsystems. The most important are:

a) \textit{Emergence} is basically described as the emergence of macroscopic behaviour from microscopic interactions of self-organised entities [\cite{holland2000emergence}]. In the context of SASO systems, this refers to the formation of patterns in the system-wide behaviour, for instance. Examples for quantification methods are [\cite{mnif2011quantitative}] and [\cite{fernandez2014information}]. In the context of quantifying life, Fernandez~[\cite{fernandez2013}] introduces the \textit{Autopoiesis}-measure which is based on emergence.

b) \textit{Self-organisation} can be expressed as a degree to which the autonomous subsystems forming an overall SASO system decide about the system's structure without external control, where the structure is expressed as interaction/cooperation/relation among individual subsystems [\cite{MST17}]. Examples of quantification methods which are using a static approach can be found in [\cite{SchmeckMCMR10}]. For methods using a dynamic approach, see [\cite{TomfordeKS17}]. An alternative discussion of self-organisation and its relation to emergence is given by [\cite{de2004emergence}].

c) \textit{Scalability} is a property that defines how far the underlying mechanisms are still promising if the number of participants grows strongly. Quantitatively, this can be measured as the impact on the control overhead, for instance.

d) \textit{Variability} or \textit{Heterogeneity} are terms referring to populations of individual subsystems as they focus on the differences in the behaviour, the capabilities or the strategies followed by the subsystems. Examples can be found in [\cite{SchmeckMCMR10}] and [\cite{lewis2015static}].

e) \textit{Mutual influences} among distributed autonomous subsystems indicate that the decisions of one have an impact (e.g., on the degree of utility achievement) of another subsystem [\cite{RudolphTH19}]. An example for a quantification technique based on the utilisation of dependency measures can be found in [\cite{RudolphHTH16}].

f) \textit{Self-adaptation} refers to the ability of systems to change their behaviour according to environmental conditions, typically to increase a utility function. A static approach to define the degree of adaptivity is given in~[\cite{SchmeckMCMR10}]. Dynamic approaches are presented in the following section.

\subsection{Quantifying the Adaptation Behaviour}
A crucial part of a lifelike system is the adaptation to its environment. The mechanism behind this self-adaptation are usually hidden from the user. When users observe a lifelike system for some time they might develop a notion of 'the normal (adaptation-)behaviour' of the system. Now, when the users note a deviation from the past behaviour the question may arise whether this deviation is the result of an abnormal process or can still be seen as normal. 

To answer that question, we need measurements to identify abnormal adaptation processes in the system. In previous work, we presented a few measures for that purpose: Configuration stability~[\cite{GollerT20}], configuration variability~[\cite{GollerTomfordeVariabilty2021}], configuration coherence~[\cite{DASC21}] and parameter utilisation~[\cite{alpaca2022}]. 

These four measures are based on the idea that the internals of the self-adaptation is not observable and therefore can be considered as a random process. This process creates a configuration vector which the PS then works with. Thus, the basis for the measures is the time series of the configurations of the system.  The configuration stability can be applied to heterogeneous systems. The other three measures require that all configurations have the same dimension and meaning for the entries.
The purpose of these measures is to detect changes over time. Therefore, the actual values are rather unimportant. The important aspects are the time series of the values and notable changes within those time series.
For following definitions let $t$ be a fixed point in time and let $S$ denote the set of all subsystems. Furthermore, we will denote a single subsystem as an agent.

\subsubsection{Configuration Stability}
\label{sec:stability}

The \textit{configuration stability measure} computes the variance of the number of active subsystems. To define 'active' in this case, we take the configuration in the current and in a previous time window, consider them as variables of a random process, then create a probabilistic density from that (i.e., following a Parzen window approach~[\cite{parzen}]), compare those densities (i.e., using the Kullback-Leibler divergence~[\cite{Bis11}]) and call an agent 'active' if the densities differ by more than a given threshold $\varepsilon$. Following an idea of Kinoshita~[\cite{kinoshita}], with $N$ being the total number of agents and $n_t$ the number of active agents, we define the activity factor $z_t := \frac{2 \cdot n_t - N + 1}{2 \cdot N}$. The activity factor is more or less the fraction of active agents. With the activity factor we calculate the fluctuation $\xi_t := z_t - \frac{1}{M} \cdot \sum^{M-1}_{i=0} z_{t-i}$ for a given window size parameter $M$. Finally, the \textit{configuration stability measure} is defined as the variance of the fluctuation: $\nu_t := \frac{1}{M} \sum_{i=0}^{M-1} \xi^2_{t-i} - \left(\frac{1}{M} \sum_{i=0}^{M-1} \xi_{t-i} \right)^2$. In short: A subsystem is called stable if its configuration does not change too much over time. Peaks in the time series of the variance of fluctuation of the number of active subsystems are an indicator for an unusual adaptation behaviour. See~[\cite{GollerT20}] for more details.

\subsubsection{Configuration Variability}
\label{sec:variability}
For this measure, we assume that 'normal' adaptation processes create only small changes in the configurations. Now, if we do a clustering of the configuration vectors before and after the adaptation, we might see a change in the number or diameter of the resulting clusters. If such a change is prominent enough it can be an indicator for an abnormal adaptation behaviour. 

Different clustering algorithms can produce different result and also small changes in one configuration vector can create a very different result. To overcome this, the measure works with fixed numbers for cluster count and then calculates the average centroid distance. Let $k$ be the number of clusters, $X_i$ a single cluster ($ 1 \leq i \leq k$) and $C_i$ the centroid of $X_i$. First, compute the distances for every data point $p$ in $X_i$ to $C_i$, take their sum and then divide by the cluster count: $s_k := \frac{\sum_{i = 1}^k \left(  \sum_{p \in X_i} \norm{p - C_i} \right)} {k}$. Repeat this for $1 \leq k \leq k_{max}$ with $k_{max} := ceil\left(\sqrt{\left| S \right|}\right)$. Finally, we define the \textit{configuration variability} $c_v$ as the average of $s_k$ over the clustering for $k=1...k_{max}$ as 
$c_v := \sum_{k=1}^{k_{max}} s_k / k_{max}$

\subsubsection{Configuration Coherence}
\label{sec:coherence}
Here, again we assume that normal adaptations produce only small changes in the resulting configuration vectors. So, for this measure we simply compute the statistical variance of all configurations: 

Let $x_s \in \RR^n$ the configuration vector of a subsystem $s \in S$ at time point $t$. Let $v_S := \sigma_S^2 =  |S|^{-1} \cdot \sum_S \left\| x_s - \hat{x} \right\|^2$ be the statistical variance of all configuration vectors.
We then define the \textit{configuration coherence} $c_{conf}$ of the system as $ c_{conf} := \frac{1}{1 + v_S}$. The value of $C_{conf}$ is in $(0;1]$ and is $1$ iff all configuration vectors are identical. Therefore, a higher value means a higher coherence. See [\cite{DASC21}] for more details.

\subsubsection{Parameter Utilisation}
\label{sec:utilisation}
For this measure, we take the possible range $V_j := max(c_j) - min(c_j)$ of values that a single entry $j$ in the configuration vector can have. This value is either fixed at design time or the result of physical limits (e.g. a maximum speed that physically possible with the used motor). Then, we fix a time window from $t-L$ to $t$ and take the range of values $V_{j,t} := max_{t-L\rightarrow t}(c_j) - min_{t-L\rightarrow t}(c_j)$  that actually were taken by the agents in the system during that period. The \textit{global parameter usage} $U_{j,g}$ is then defined as $U_{j,g} := V_{j,t} / V_j$. 

The \textit{average parameter usage}  $U_{j,a}$ is defined as $U_{j,a} := \sum_{a_i \in A} V_{j,t, a_i} / (V_j \cdot |A|)$ where $A$ is the set of agents, $a_i$ a single agent and $V_{j,t, a_i}$ is the range of values that the configuration parameter $j$ took in the configuration of the single agent $a_i$ during the time frame. This is the average size of taken values divided by the total possible range. See [\cite{alpaca2022}] for more details.

These two measure again are based on the idea that a notable change in the range of actually taken configuration values indicates unusual behaviour.

%%%%%%%%%%%%%%%%%%%%%%%%%%%%%%%%%%%%%%%%%%%%%%%%%%%%%%%%%%%%%%%%%%%%%%%
%
%
%       Sec 3: Extented Framework
%
%
%%%%%%%%%%%%%%%%%%%%%%%%%%%%%%%%%%%%%%%%%%%%%%%%%%%%%%%%%%%%%%%%%%%%%%%

\section{III. An Extended Measurement Framework for Lifelike Computing Systems}

Considering the concept of lifelike technical systems and their desired capabilities, the set of existing metrics is probably not sufficient enough to cover the entire behaviour. In particular, we will have to investigate new techniques to measure lifelike attributes. Although there is currently no exact definition of what lifelike computing systems are, we can approach the question of what is missing in the measurement framework by considering 'qualities of life' that we aim to transfer to technical usage and that go beyond the SASO-based scalability, adaptation, organisation, or robustness questions.

\subsection{Possible Quantification Aspects}

So far, we identified the following aspects of lifelike systems as options for further investigation:

a) A lifelike system will try to adapt to the niche in which the system survives. The result of this adaptation may be expressed with a measure of 'fitness in the niche' or 'degree of the niche appropriateness'. 

b) A lifelike system will evolve over time. Consequently, another measure should aim at quantifying the evolution behaviour itself. One reason for evolution is the adaptation to niche of survival. Therefore such a measure could be related the 'fitness of the niche'. Another reason for evolution could be the change of the primary system purpose. Therefore, a second aspect of quantifying the evolution behaviour is the coverage of the primary purpose. The latter case continues the ideas formulated in the Organic Computing initiative when defining the property of 'flexibility', i.e. how far a SASO system can react appropriately to changing goal functions [\cite{BeckerHT12}].

c) An evolution process implies that the system is somehow converted (or better: converts itself). Besides the description of this process of time, a more static measure based on the design can aim at determining a 'degree of convertability', i.e. the freedom to which the system can evolve during operation.

d) Such a lifelike, evolutionary behaviour is done in the context of the environmental conditions, which includes the presence of other subsystems in open system constellations. As a result, parts of the decisions of a lifelike system are about the current integration into such a constellation, resulting in 'self-improving system integration' [\cite{BellmanBDEGLLNP21}]. Although there is currently no integration measure available, recent work suggests that such an integration state is probably a multi-objective function that builds upon metrics mentioned in the context of SASO measures [\cite{GruhlTS18}].

e) A lifelike character has implications on the way we design and operate systems. In contrast to current practices that take design-related decisions and provide corridors of freedom for the self-* mechanisms, design-time decisions themselves need to become reversible or changeable by the systems, resulting in a degree of reversibility (in a static manner) or changes (in the sense of how strong the design has already been altered).

f) We expect that at one point lifelike systems will be given tasks that require the interaction with other lifelike systems. Consequently, this will lead to groups of systems that interact with each other and even with humans. At this point the notion of 'social behaviour' will become important which implies the need for measuring social behaviour and capabilities.

%%%%%%%%%%%%%%%%%%%%%%%%%%%%%%%%%%%%%%%%%%%%%%%%%%%%%%%%%%%%%%%%%%%%%%%
%
%
%       Sec 3: Transferability 
%
%
%%%%%%%%%%%%%%%%%%%%%%%%%%%%%%%%%%%%%%%%%%%%%%%%%%%%%%%%%%%%%%%%%%%%%%%

\section{IV. A Measure for Quantifying Transferability}

The niche of survival of a SASO system is defined by two parts: The actual environment the SASO system operates in and the goal the system is given. If the system experiences extreme and sudden changes in the environment or is given a new system goal is has to deal with a new niche of survival. In this section, we present an approach to measure the transferability of a lifelike system from one niche to another. 

Taking a look into the real world, we might find several possible definitions of transferability. One that is fast at hands is: A system that only depends loosely on its environment can be transferred more easily to a new environment than a system whose performance is heavily influenced by the environment.

Fernandez gives an example~[\cite{fernandez2013}]: Bacteria can reproduce mostly by themselves. Viruses on the other hand are highly dependent on their hosts to reproduce. 
Therefore, it seems evident to define transferability by some kind of relationship between the environment and the system states. Since the system itself changes over time due to self-adaptation and self-organisation we can state that the notion of transferability is not static. 

While active a SASO system observes its environment and its inner states and then will create a new inner state (that might be identical to the previous one). All those states can be seen as information, at the latest after they are transformed into a suitable representation for the necessary computations in the hardware. Therefore, we can assign information theoretic measures to them such as emergence or complexity. 

For the following, we will use the definitions for these measures as given by~[\cite{fernandez2013}]: For an information represented as a string $X=x_0x_1...$ over an finite alphabet $A$ of length $n$ the \textit{emergence} $E$ is defined as the normalised information entropy $I$ with: $E := I = -K \sum_{i=1}^n p_i~log~p_i$ where $p_i$ is the probability of the symbol $x_i$ occurring in $X$ and K the normalisation factor. The \textit{complexity} $C$ is then defined by $C:= 4\cdot E \cdot (1-E)$.

Now, we can define the \textit{transferability measure} as the (Pearson-)correlation between the complexity of the environment and the complexity of the system's configuration (as a representation for the inner states). It is computed for a given point in time and with a limited list of past values for the complexity:

Let $X_i$ be the system states at time point $i$ and let $Y_i$ be those of the environment. Let $t$ be a fixed point in time and let $X = \{X_j\}$ and $Y = \{Y_j\}$ for $t-l \geq j \geq t$ be the set of states in a given time window of size $l$.

Let $c:=Corr(X,Y)$ be the correlation coefficient of $X$ and $Y$. We then define the \textit{transferability} of the system as
\begin{equation}
T := 1 - \left|c\right|
\end{equation}
So, when X and Y have a low correlation and therefore are mostly independent from each other, the transferability will take a higher value. This reflects the idea that systems that are mostly independent of their environment are easier to transfer to another niche of survival.

%%%%%%%%%%%%%%%%%%%%%%%%%%%%%%%%%%%%%%%%%%%%%%%%%%%%%%%%%%%%%%%%%%%%%%%
%
%
%       Sec 4: Eval
%
%
%%%%%%%%%%%%%%%%%%%%%%%%%%%%%%%%%%%%%%%%%%%%%%%%%%%%%%%%%%%%%%%%%%%%%%%
\section{V. Evaluation}
In this section, we will apply the fore-mentioned measures to three simulation scenarios. The first scenario is an artificial traffic simulation as an example of a system that is heavily influenced by its environment. The second scenario is an artificial flocking~[\cite{flocking}] simulation. This system is an example of a self-contained system without any real environment. Finally, we will take a look at a cellular automaton in the form of a "Game of Life"-Simulation~[\cite{games1970fantastic}] which is prototypical for the analysis of emergence and complexity. All simulations are implemented in the MASON simulation environment~[\cite{mason}].

For the evaluation of the transferability measure, we chose for the window size a value of $L=40$. Our Experiments show that lower values give noisier graphs and short term effects have can have an unwanted impact on the result. Higher values give smoother graphs which give better insight in the long term behaviour but tend to cover relevant changes. Nevertheless, the best choice for $L$ depends on the investigated system.

%%%%%%%%%%%%%%%%%%%%%%%%%%%%%%%%%%%%%%%%%%%%%%%%%%%%%%%%%%%%%%%%%%%%%%%
%
%
%       Eval: Traffic
%
%
%%%%%%%%%%%%%%%%%%%%%%%%%%%%%%%%%%%%%%%%%%%%%%%%%%%%%%%%%%%%%%%%%%%%%%%
\subsection{Traffic simulation}

This scenario is inspired by the Organic Traffic Control (OTC) system that self-adapts the green duration of traffic lights and learns the best adaptation strategy over time~[\cite{ProthmannRTBMS08}]. 
Here, we used an abstract traffic simulation, where the SASO system consists of several interconnected intersection controllers. 

Each intersection tries to minimise the waiting time for all cars at all incoming lanes by optimising the green/red light times for each lane. The intersections do not communicate with each other. The cars in the simulation are part of the environment. They pick their destination randomly and chose the shortest path to it. When they reach their destination, the process is repeated. 

\begin{figure}[hbt!]
    \centering
    \includegraphics[scale=0.45]{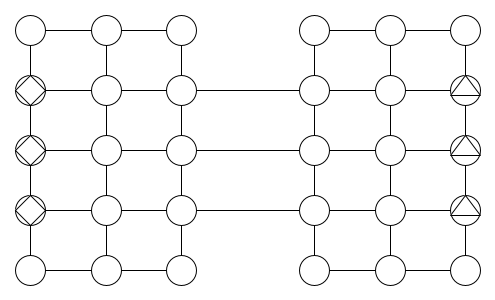}
    \caption{The street network for the second traffic simulation. Circles are intersections and the lines represent the streets. Home steads are marked with a square, workplaces with a triangle}
    \label{fig:rushhour_network}
\end{figure}

Our scenario simulates rush hours in a city street network with two islands each with a Manhattan-type network of sizes 3 by 5. The islands are connected with three bridges. The connections between two intersections provide one lane for each direction (see Figure~\ref{fig:rushhour_network}]. 

During the whole simulation, 250 cars driving around randomly. Furthermore, three intersections on one island are designated to be homesteads and three intersections on the other island are labelled as workplaces. At $t=250$ a total count of 500 new cars will appear randomly in the three homesteads and start driving towards one of the three workplaces -- which models a disturbance in the system. Therefore, the bridges have to handle an increasing flow from one side to the other in the following time steps. When the new cars arrive at their workplaces, they are removed from the simulation. At $t=750$ this is repeated in the opposite direction.

\begin{figure}[hbt!]
	\centering
    \includegraphics[width=1\linewidth]{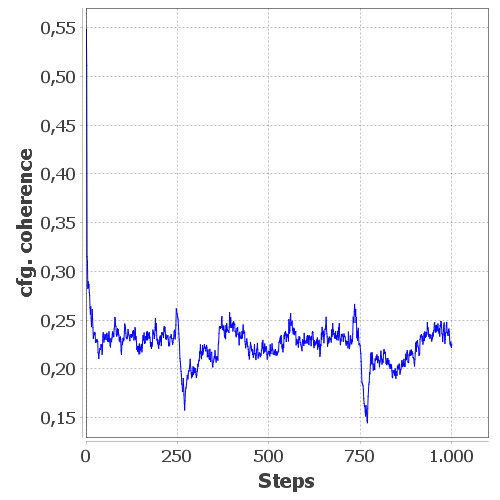}
    \caption{The time series of the configuration coherence for the traffic simulation.}
    \label{fig:rushhour_cfg_coherence}
\end{figure}

\begin{figure}[hbt!]
	\centering
	\includegraphics[width=1\linewidth]{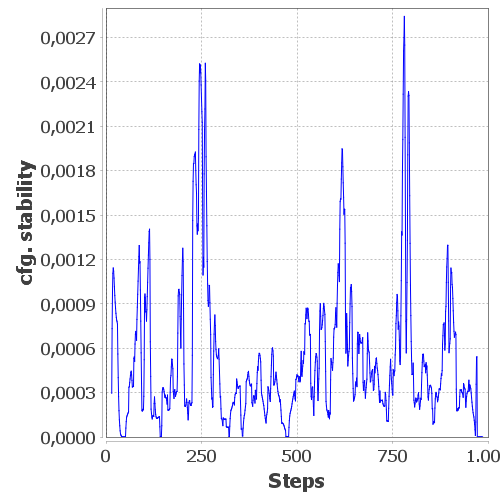} 
     \caption{The configuration stability of the traffic simulation. $M=L=15$, $\varepsilon=2$}
    \label{fig:rushhour_cfg_stability}
\end{figure}

\begin{figure}[hbt!]
	\centering
	\includegraphics[width=1\linewidth]{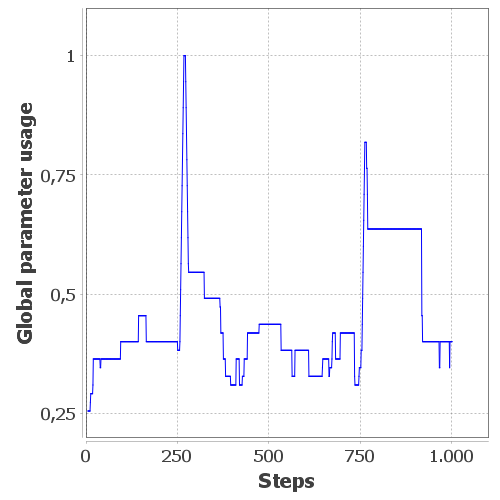} 
    \caption{The global parameter usage for the traffic simulation. $L=5$}
    \label{fig:rushhour_global_usage}
\end{figure}

\begin{figure}[hbt!]
	\centering
	\includegraphics[width=1\linewidth]{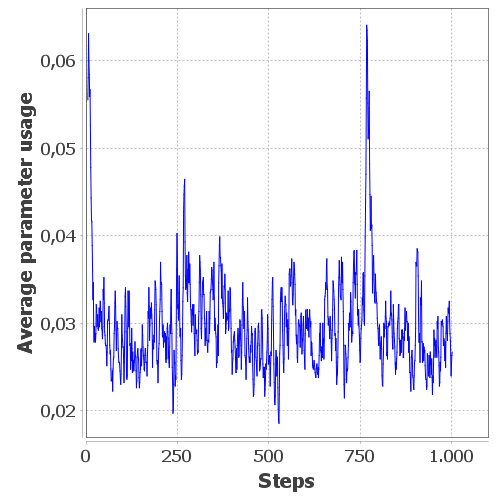} 
    \caption{The average parameter usage for the traffic simulation. $L=5$}
    \label{fig:rushhour_average_usage}
\end{figure}

\begin{figure}[hbt!]
	\centering
	\includegraphics[width=1\linewidth]{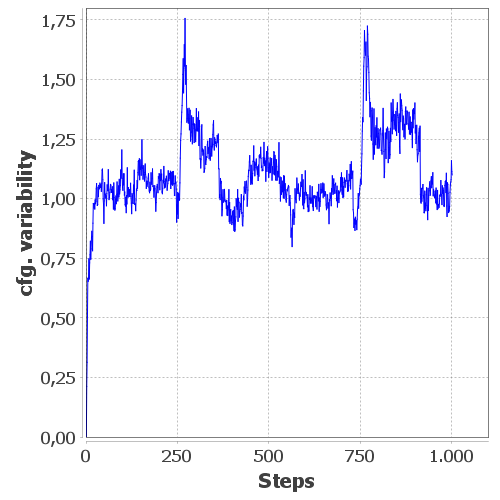} 
    \caption{The configuration variability for the traffic simulation}
    \label{fig:traffic_variability}
\end{figure}

Figures~\ref{fig:rushhour_cfg_coherence}, \ref{fig:rushhour_cfg_stability}, \ref{fig:traffic_variability}, \ref{fig:rushhour_average_usage} and \ref{fig:rushhour_global_usage} show the results for this simulation. In all five graphs the impact of the increased traffic flow is visible. The configuration stability on its own is prone to the random background noise in the simulation. It shows several other prominent peaks. The average parameter usage fails to identify the morning rush hour. The corresponding peak is not high enough. But together the measures give a good indication for the two unusual events.

For the transferability measure we need to define the sources of information for the environment and the internal states. For the environment, we take the current number of cars waiting on the incomming lanes for each crossing. The internal states are the red light times for those lanes. Both values are discrete (the latter by system design), so we can easily calculate the probabilistic distribution for the complexity.

\begin{figure}[hbt!]
	\centering
	\includegraphics[width=1\linewidth]{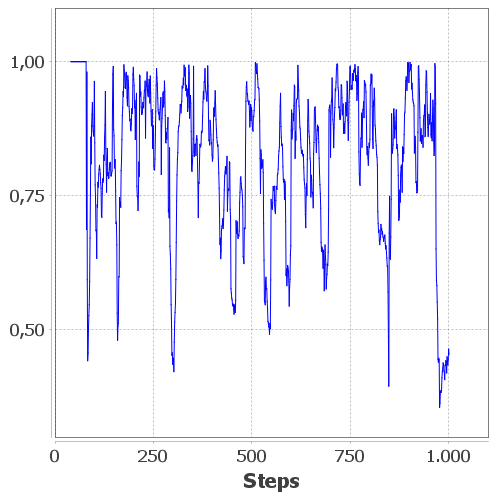}
	\caption{Transferability of the rushhour simulation. $L=40$, average value = 0.81}
	\label{fig:rushour_transferability}
\end{figure}

Figure~\ref{fig:rushour_transferability} shows the result of the transferability measure. With an average value of 0.81 and nearly all values above 0.5 we can say that the system is fairly transferable. That becomes apparent if we transfer the system to an environment where it no longer optimises traffic flows for cars but any other kind of flow: Liquids, gases or maybe communication packets in a network.

%%%%%%%%%%%%%%%%%%%%%%%%%%%%%%%%%%%%%%%%%%%%%%%%%%%%%%%%%%%%%%%%%%%%%%%
%
%
%       Eval: Flocking
%
%
%%%%%%%%%%%%%%%%%%%%%%%%%%%%%%%%%%%%%%%%%%%%%%%%%%%%%%%%%%%%%%%%%%%%%%%

\subsection{Flocking simulation}

The flocking simulation consists of 50 birds on a toroidal plane with random starting points and random initial orientations. All birds follow the usual rules of flocking: \begin{itemize}
    \item Alignment: A bird will align its direction with the average direction of its neighbours
    \item Cohesion: A bird will steer towards the centre of all neighbouring birds
    \item Avoidance: A bird will steer away from neighbours that are too close
\end{itemize}
For each of these rules, a direction vector is computed, weighted with a factor and then added together. The result is normalised and then added to the current direction vector.

\begin{figure}[hbt!]
    \centering
    \includegraphics[width=.66\linewidth]{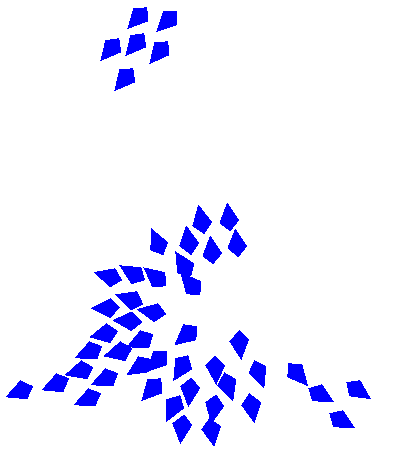}
    \caption{The flocking simulation at $t=503$. The shot occurred near the centre of the lower cluster.}
    \label{fig:bird_shot}
\end{figure}

We introduce a disturbance as the main trigger for adaptations as follows: At time point $t=500$ one bird $A_0$ is being shot at. All birds within a distance of 50 units fly diametrically away from $A_0$ for two time steps and then follow their usual behaviour again. Figure~\ref{fig:bird_shot} shows the simulation at $t=503$ with the result of the shot.

\begin{figure}[hbt!]
	\centering
    \includegraphics[width=1\linewidth]{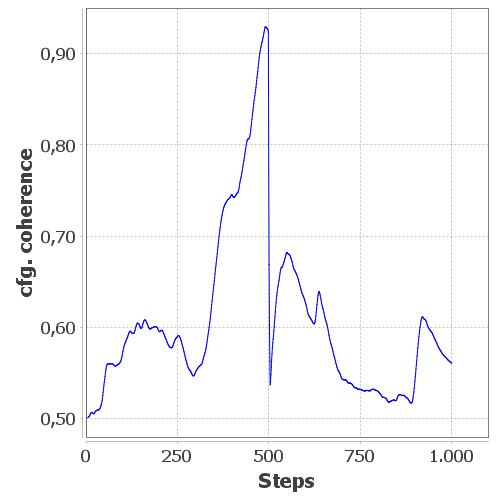}
    \caption{The time series of the configuration coherence for the flocking simulation.}
    \label{fig:flock_coherence}
\end{figure}

\begin{figure}[hbt!]
	\centering
	\includegraphics[width=1\linewidth]{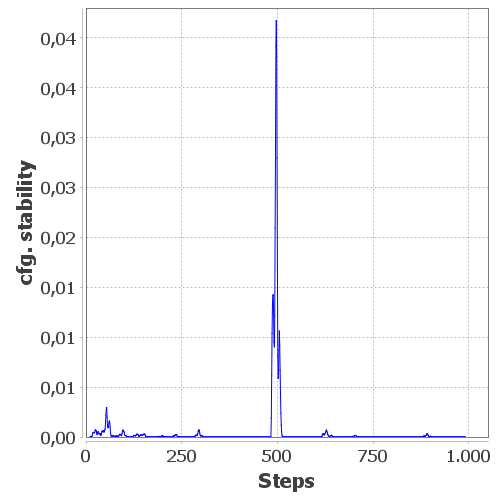} 
     \caption{The configuration stability of the flocking simulation. $M=L=10$, $\varepsilon=1$}
    \label{fig:flock_stability}
\end{figure}

\begin{figure}[hbt!]
	\centering
	\includegraphics[width=1\linewidth]{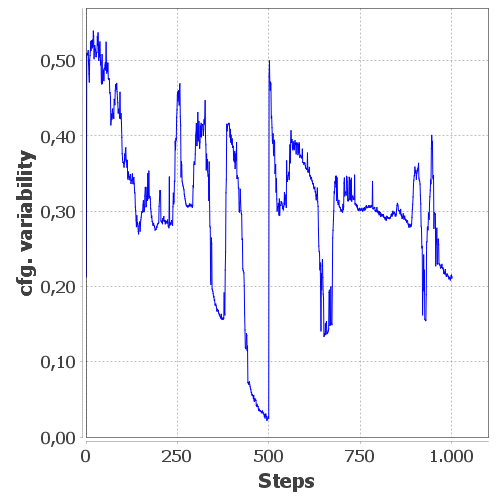} 
    \caption{The configuration variability for the flocking simulation}
    \label{fig:flock_variability}
\end{figure}

\begin{figure}[hbt!]
	\centering
	\includegraphics[width=1\linewidth]{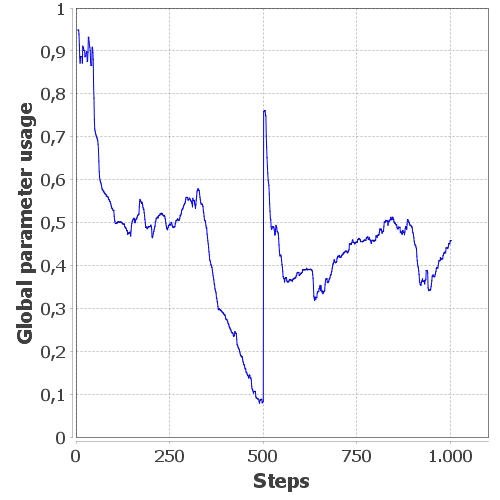} 
    \caption{The global parameter usage for the flocking simulation. $L=5$}
    \label{fig:flock_global_usage}
\end{figure}

\begin{figure}[hbt!]
	\centering
	\includegraphics[width=1\linewidth]{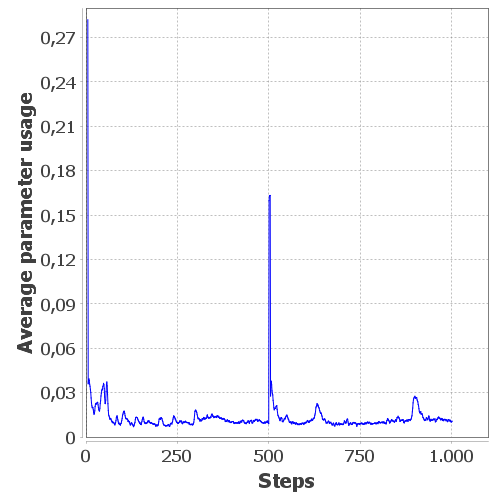} 
    \caption{The average parameter usage for the flocking simulation. $L=5$}
    \label{fig:flock_average_usage}
\end{figure}

Figures~\ref{fig:flock_coherence}, \ref{fig:flock_stability}, \ref{fig:flock_variability}, \ref{fig:flock_global_usage} and \ref{fig:flock_average_usage} show the result of the adaptation assessment measures. All of them show the impact of the disturbance. The configuration variability on its own would give several additional signals. The configuration stability and the average parameter usage on the other hand are very sensitive for this disturbance.  

\begin{figure}[hbt!]
	\centering
	%
	% Buckets = 100, window = 40, offset = 1
	%
	\includegraphics[width=1\linewidth]{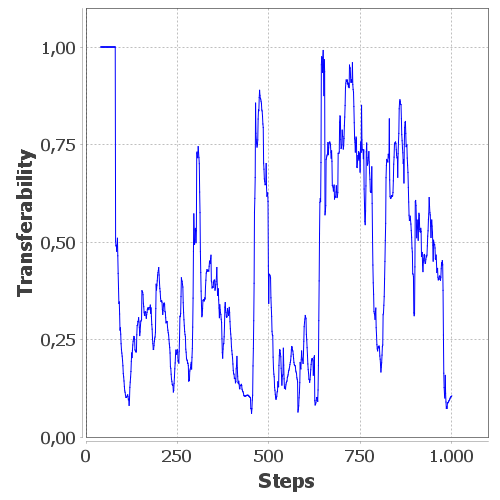}
	\caption{Transferability of the flocking simulation. $L=40$, average value = 0.43}
	\label{fig:flock_transfer}
\end{figure}

Finally, figure~\ref{fig:flock_transfer} shows the time series of the transferability. Since entropy is defined on discrete distributions, we decided to discretize the angles of the birds using a histogram with 100 buckets. Extension to the continuous case remains as future work.

The flocking rules imply that the next state of an agent heavily depends on the current states of the neighbours. Phases of a low transferability correspond with phases of a wide range of used angles. Phases of fewer used angles (meaning bigger clusters) come with a higher transferability. The average transferability value over the whole simulation is 0.43. Note that the impact of the shot is visible at $t=460$, the beginning of the frames that include the time point of the disturbance.

%%%%%%%%%%%%%%%%%%%%%%%%%%%%%%%%%%%%%%%%%%%%%%%%%%%%%%%%%%%%%%%%%%%%%%%
%
%
%       Eval: Game of life
%
%
%%%%%%%%%%%%%%%%%%%%%%%%%%%%%%%%%%%%%%%%%%%%%%%%%%%%%%%%%%%%%%%%%%%%%%%

\subsection{Game of life}
\begin{figure}[hbt!]
	\centering
    \includegraphics[width=1\linewidth]{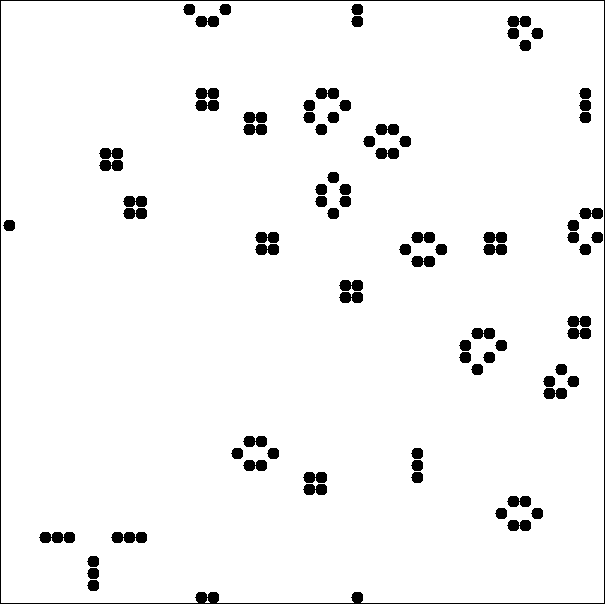}
    \caption{The state of the Game of Life simulation at $t=302$.}
    \label{fig:gol_state}
\end{figure}

This Game of Life simulation is set on a toroidal plane with 50x50 cells.
Each cell has one of two states: 'dead' or 'alive' (represented as 0 or 1). The state is the configuration we are observing. Each cell computes the state for the next step using the 'classical' rules depending on the number of living neighbours:
\begin{itemize}
    \item It dies if fewer than two or more than three neighbours live 
    \item It stays alive if two or three neighbours live
    \item It becomes alive if exactly three live neighbours are alive.
\end{itemize}
After a random initialisation, the simulation reaches a state with constant patterns and a few 'blinkers' after 300 steps. Figure~\ref{fig:gol_state} shows the simulation after 302 steps

\begin{figure}[hbt!]
	\centering
    \includegraphics[width=1\linewidth]{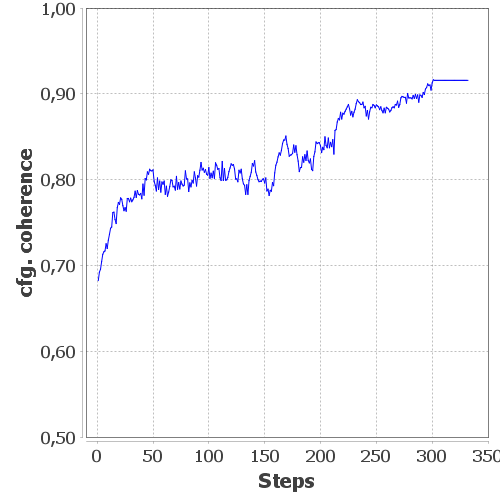}
    \caption{The time series of the configuration coherence for the Game of Life simulation.}
    \label{fig:gol_coherence}
\end{figure}

\begin{figure}[hbt!]
	\centering
	\includegraphics[width=1\linewidth]{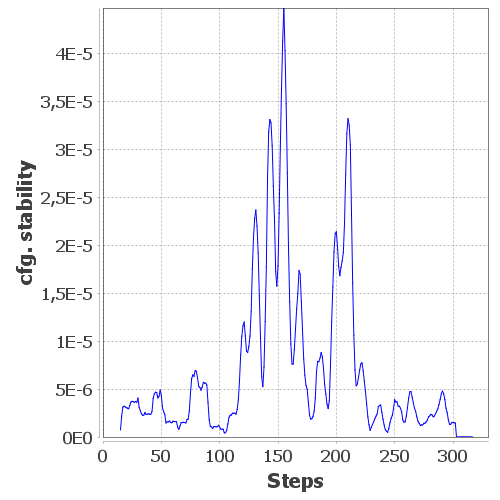} 
     \caption{The configuration stability of the Game of Life simulation. $M=L=15$, $\varepsilon=0.05$}
    \label{fig:gol_stability}
\end{figure}

\begin{figure}[hbt!]
	\centering
	\includegraphics[width=1\linewidth]{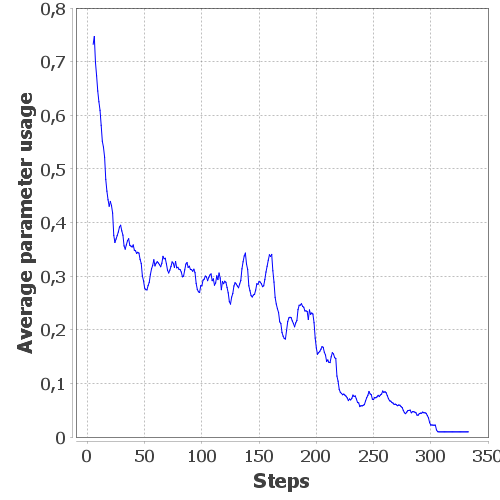} 
     \caption{The average parameter usage of the Game of Life simulation. $L=5$}
    \label{fig:gol_avg_parameter_usage}
\end{figure}

\begin{figure}[hbt!]
	\centering
	\includegraphics[width=1\linewidth]{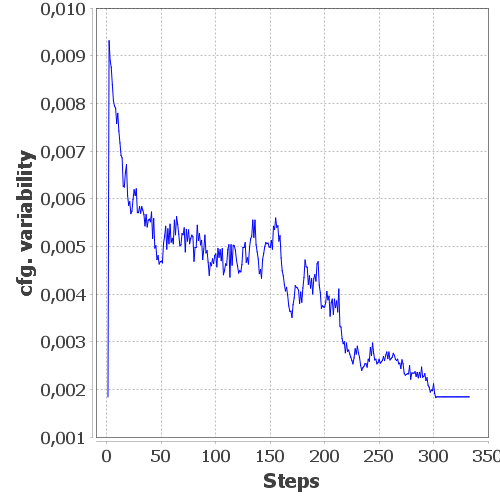} 
    \caption{The configuration variability for the Game of Life simulation}
    \label{fig:gol_variability}
\end{figure}

The number of changing cells decreases over time. Therefore, the configuration coherence increases, see figure~\ref{fig:gol_coherence}. Since there are still changing cells in the end the graph does not reach 1. For the same reasons the configuration variability (see figure~\ref{fig:gol_variability}) and the average parameter usage (see figure~\ref{fig:gol_avg_parameter_usage} drop but do not reach 0. Figure~\ref{fig:gol_stability} shows the configuration stability of the simulation. The higher values from $t=120$ to $t=220$ are the result of a high activity with lots of changing patterns. Since there are dead and alive cells present at all times the global range of taken values is the same as the possible range. Therefore, the global parameter usage for this simulation is always equal to $1$.

This simulation again is self-contained. It defines its own environment. The average transferability value is $0.41$ and shows phases of high correlation in the complexity of the current and the previous states and other phases with low correlation. This is a result of the nature of the simulation and its changing complexity.

\begin{figure}[hbt!]
	\centering
	%
	% Buckets = 5 (obwohl binär-werte kommt mit 2 irgendwie müll raus...)
	% window = 40, offset = 1
	%
	\includegraphics[width=1\linewidth]{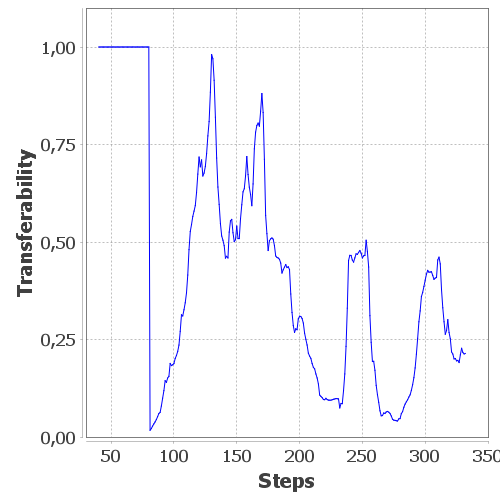}
	\caption{Transferability of the flocking simulation. $L=40$, average value = 0.41}
	\label{fig:gol_transfer}
\end{figure}

\subsection{Evaluation summary}
The evaluations of the measures for the adaptation behaviour show that they are, if used together, capable of identifying unusual adaptation processes in the first two simulations. In the third simulation, they reflect the system behaviour with it's decreasing activity and increasing order. 

In all three simulation, the transferability measure shows the relationship between the environment and the system. In the traffic scenario the measure shows that the system mostly reacts to the environment and does not depend too much on its inner states. The other two scenarios show that system that depend on their own states as well as the one of their environment lead to a lower transferability. 

\section{VI. Conclusions}

In this paper, we outlined our idea of lifelike systems as an extension of SASO systems with increased autonomy. Based on this, we reviewed approaches to quantify system behaviour mainly at the macro-level. Here, we focused on measures for self-adaptation. 
Thereafter, we discussed possible approaches for an extended measurement framework for lifelike systems. As a first step toward such a framework, we presented a measure for the transferability of lifelike systems. 

In our opinion, an important property of lifelike systems will be the ability for self-explanation of its observable behaviour and the underlying decisions. Complex, evolving systems with high autonomy will most likely face acceptance problems if they are unable to present explanations for their actions. Self-explanation has to answer two major questions: i) When to provide self-explanations to the user and ii) what is explained (including 'how'). This paper proposed to address the first question by using a measurement framework and gave a few possible measures for this.

Future work will address measurements to quantify other aspects of lifelike systems such as the evolution process or social behaviour. Following this, the final goal of this research is to provide mechanisms and techniques that eventually provide human-understandable self-explanations.

\footnotesize
\bibliographystyle{apalike}
\bibliography{references}

\end{document}